\documentclass[preprint,preprintnumbers,amsmath,amssymb]{revtex4}

\usepackage{bm}
\usepackage{mathrsfs}

\begin{document}


\title{Infinite-Hamiltonian Structures of the Riemann Equation}

\author{Refik Turhan}
\email{refik@eng.ankara.edu.tr}
\affiliation{Department of Engineering Physics, Faculty of Engineering, 
Ankara University, 06500 Ankara, Turkey.}

\date{\today}

\begin{abstract}
We give two distinct infinite-Hamiltonian representations for the Riemann equation. One with first order Hamiltonian operators and another with third order-first order Hamiltonian operators. Both representations contain an arbitrary function of the dynamical variable. We prove that all the Hamiltonian operators of the Riemann equation are mutually compatible.
\end{abstract}

\maketitle

\section{Introduction}

The standard bi-Hamiltonian theory of integrability requires an evolution equation to be a member of a hierarchy of equations all of which can be written in Hamiltonian form by two distinct compatible Hamiltonian operators in the same set of dependent and independent variables \cite{Mag,GD}. I.e if ${\mathcal H}$ and ${\mathcal K}$ are two Hamiltonian (skew-adjoint, Jacobi identity satisfying) operators which are compatible; meaning that ${\mathcal H} + \alpha {\mathcal K}$ is also Hamiltonian for any $\alpha \in {\mathbb{R}}$, then the hierarchy of  evolution equations
\begin{equation}
u_{t_{n}}=F_{n}[u]={\mathcal H}{\rm E}(H_{n})={\mathcal K}{\rm E}(H_{n-1}),\;\;\; n=0,1,2,3,\cdots \nonumber
\end{equation}
are completely (bi-Hamiltonian) integrable provided (technically) that for all $n$, $H_{n}$ exits. Here, $H_{n}$ are the densities of the Hamiltonian functionals ${\rm H}_{n}=\int H_{n}{\rm d}x$, ${\rm E}$ is the Euler or functional derivative and $u=u(t_{n},x)$ is the dynamical variable dependent on a spatial variable $x$ and evolution parameters  $t_{n}$. 

Almost all of the known bi-Hamiltonian hierarchies conform to the above given standart description of the theory. However, there are exceptional examples of hierarchies where the Hamiltonian operators (by which the hierarchies are written) are reported to be incompatible \cite{OE,ON,OL}.

The bi-Hamiltonian integrability is the minimum in terms of the number of distinct Hamiltonian operators involved in writing the hierarchies. There are examples of multi-Hamiltonian hierarchies where the equations are written in Hamiltonian form by more than two distinct Hamiltonian operators with both nontrivial \cite{ON,NUT1,ANNOV,GN} and trivially related Hamiltonian operators \cite{BK,AF2,AF3,ENG}. 

The Riemann equation
\begin{equation}
u_{t}=uu_{x} \label{RIE}
\end{equation}
occupies a special place in the literature because of both the number of Hamiltonian representations given for it and as an example of equations some of whose Hamiltonian operators are reported to be incompatible. There are four distinct Hamiltonian representations given for it with the Hamiltonian operators \cite{ON}
\begin{equation} 
{\mathcal K}_{0}=D,\;\;{\mathcal K}_{1}=\sqrt{u}D\sqrt{u},\;\;{\mathcal K}_{2}=uDu,\
\;\;{\mathcal K}_{3}=D\frac{1}{u_{x}}D\frac{1}{u_{x}}D, \nonumber
\end{equation}
where $D^{n}={\rm d}^{n}/{\rm d}x^{n}$, $n \in {\mathbb {N}}$.
It is mentioned, however, that the operator ${\mathcal K}_{3}$ is compatible only with ${\mathcal K}_{0}$ but not with the other two operators ${\mathcal K}_{1}$ and ${\mathcal K}_{2}$ \cite{ON}.
 
In the present work we show explicitly that the Riemann equation can be written in Hamiltonian form by first order Hamiltonian operators in infinitely many different ways. And also by third order-first order compatible Hamiltonian operators again in infinitely many different ways. In the Hamiltonian representations given here, both the Hamiltonian operators and the corresponding hierarchies of symmetries and densities contain arbitrary functions of $u$. 
We also show that all of the newly given Hamiltonian operators and the ones given earlier are compatible operators.

\section{Hierarchies From the First Order Hamiltonian Operators}

It is well known \cite{GD} that for any arbitrary function $f$ of $u$, 
\begin{equation}
{\mathcal J}_{f}=fDf=f^{2}D+ff_{u}u_{x} \label{GDO}
\end{equation}
is a Hamiltonian operator. Obviously, any two such operators ${\mathcal J}_{f}$ and ${\mathcal J}_{g}$ are compatible Hamiltonian operators. Irrespective of what the functions $f$ and $g$ explicitly are, the following two way hierarchy can be easily constructed by the operators ${\mathcal J}_{f}$ and ${\mathcal J}_{g}$.
\begin{eqnarray}
\begin{array}{lccr}
&&\vdots& \\
u_{t_{-2}}=&\frac{3f^{3}}{2g^{2}}(\frac{f}{g})_{u}u_{x}&=\
{\mathcal J}_{g}{\rm E}(\int{\frac{3f^{4}}{8g^{5}}{\rm d}u})=&\
{\mathcal J}_{f}{\rm E}(\int{\frac{f^{2}}{2g^{3}}{\rm d}u}) \\
u_{t_{-1}}=&f(\frac{f}{g})_{u}u_{x}&=\
{\mathcal J}_{g}{\rm E}(\int{\frac{f^{2}}{2g^{3}}{\rm d}u})=&\
{\mathcal J}_{f}{\rm E}(\int{\frac{1}{g}{\rm d}u}) \\
u_{t_{0}}=&0&={\mathcal J}_{g}{\rm E}(\int{\frac{1}{g}{\rm d}u})=&\
{\mathcal J}_{f}{\rm E}(\int{\frac{1}{f}{\rm d}u})\\
u_{t_{1}}=&g(\frac{g}{f})_{u}u_{x}&=\
{\mathcal J}_{g}{\rm E}(\int{\frac{1}{f}{\rm d}u})=&\
{\mathcal J}_{f}{\rm E}(\int{\frac{g^{2}}{2f^{3}}{\rm d}u})\\
u_{t_{2}}=&\frac{3g^{3}}{2f^{2}}(\frac{g}{f})_{u}u_{x}&=\
{\mathcal J}_{g}{\rm E}(\int{\frac{g^{2}}{2f^{3}}{\rm d}u})=&\
{\mathcal J}_{f}{\rm E}(\int{\frac{3g^{4}}{8f^{5}}{\rm d}u})\\
&&\vdots&
\end{array}\label{HIE1}
\end{eqnarray}

Any third Hamiltonian operator  ${\mathcal J}_{h}$ of form  (\ref{GDO}) can be trivially related to the operators ${\mathcal J}_{g}$ and  ${\mathcal J}_{f}$ as
\begin{equation}
{\mathcal J}_{h}= {\mathcal J}_{f}{\mathcal J}_{g}^{-1}{\mathcal J}_{f}, \nonumber
 \end{equation}
only when $f/g=constant$. In case of which, however,
the operators ${\mathcal J}_{h}$ and ${\mathcal J}_{f}$ are  constant multiples of the operator ${\mathcal J}_{g}$. Therefore, each specific pair of unequal functions $f$ and $g$ generate  formally a different hierarchy.

All of the members of the symmetry hierarchy (\ref{HIE1}) above are first order partial differential equations (PDEs). Since all first order PDEs are transformable to each other up to the contact transformations \cite{IBR}. One can, as a rule, take any specific one of the symmetries and transform it (and accordingly the other quantities) to the Riemann equation in the form (\ref{RIE}) or to any desired other first order equation.  Thus, the above hierarchy can viewed as a Riemann hierarchy which is written in a rather superficially  general form: two arbitrary functions in each of the arbitrary coefficients of the $u_{x}$ terms in the symmetry hierarchy. 
Therefore, we proceed with relating the arbitrary functions $f$ and $g$ in such a way that the symmetry $u_{t_{1}}=g(g/f)_{u}u_{x}$ in (\ref{HIE1}) is the Riemann equation in the form (\ref{RIE}). I.e we restrict 
\begin{equation}
g(\frac{g}{f})_{u}=u.\label{CHO}
 \end{equation}
Then, in terms of a new arbitrary function $r=r(u)$, introduction of which is only for convenience, $f$ and $g$ becomes $f=u/rr_{u}$ and $g=u/r_{u}$. Consequently, ${\mathcal J}_{f}$ and ${\mathcal J}_{g}$ reduce to
\begin{equation}
\mathcal{J}_{1}=\frac{u}{rr_{u}}D\frac{u}{rr_{u}},\;\;\;\;\mathcal{J}_{2}=\
\frac{u}{r_{u}}D\frac{u}{r_{u}},\nonumber
\end{equation}
respectively. Here $rr_{u} \neq 0$ is assumed. As a result, we get the following 
hierarchy
\begin{eqnarray}
\begin{array}{lccr}
&&\vdots& \\
u_{t_{-3}}=&-\frac{3}{2}r^{-5}uu_{x}&=\
{\mathcal J}_{2}{\rm E}(\int{\frac{3r^{-4}r_{u}}{8u}{\rm d}u})=&\
{\mathcal J}_{1}{\rm E}(\int{\frac{r^{-2}r_{u}}{2u}{\rm d}u}) \\
u_{t_{-2}}=&-r^{-3}uu_{x}&=\
{\mathcal J}_{2}{\rm E}(\int{\frac{r^{-2}r_{u}}{2u}{\rm d}u})=&\
{\mathcal J}_{1}{\rm E}(\int{\frac{r_{u}}{u}{\rm d}u}) \\
u_{t_{-1}}=&0&={\mathcal J}_{2}{\rm E}(\int{\frac{r_{u}}{u}{\rm d}u})=&\
{\mathcal J}_{1}{\rm E}(\int{\frac{rr_{u}}{u}{\rm d}u}) \\
u_{t_{0}}=&uu_{x}&={\mathcal J}_{2}{\rm E}(\int{\frac{rr_{u}}{u}{\rm d}u})=&\
{\mathcal J}_{1}{\rm E}(\int{\frac{r^{3}r_{u}}{2u}{\rm d}u}) \\
u_{t_{1}}=&\frac{3}{2}r^{2}uu_{x}&={\mathcal J}_{2}{\rm E}\
(\int{\frac{r^{3}r_{u}}{2u}{\rm d}u})=&\
{\mathcal J}_{1}{\rm E}(\int{\frac{3r^{5}r_{u}}{8u}{\rm d}u}) \\
&&\vdots&
\end{array} \nonumber
\end{eqnarray}
It is clear from the above hierarchy that the Riemann equation can be written in Hamiltonian form in infinitely many  ways because of the arbitrariness 
of the function $r$. Associated with each specific $r$ there exist a hierarchy of Hamiltonian densities in involution. Therefore, such zeroth order (in $x$-derivatives of $u$) hierarchies of Hamiltonian densities are infinitely many. In other words, the Riemann equation admits any desired function of $u$ as a zeroth order density. A fact which was noted in \cite {ON} even though only the polynomial zeroth order densities of the Riemann equation were explicitly written in the multi-Hamiltonian hierarchies given there.

Obviously, the above choice of the symmetry to be restricted is only one of the other infinitely many possibilities. One can consider another symmetry in the hierarchy (\ref{HIE1}), take its $u$ dependent arbitrary coefficient, restrict it to $u$ and do the rest as above. The result is again a Riemann hierarchy containing the Riemann equation in the form (\ref{RIE}).
 
Moreover, the above construction can also be used to directly obtain the infinite-Hamiltonian representations of the first order PDEs other than the Riemann equation.  For example, if instead of the restriction (\ref{CHO}), we were chosen 
\begin{equation}
g(\frac{g}{f})_{u}=1, \nonumber
\end{equation}
we would directly arrive at the infinite-Hamiltonian hierarchy of the equation
\begin{equation}
u_{t}=u_{x}.\nonumber
\end{equation}

The above construction is not the only construction where the Riemann equation is infinite-Hamiltonian. There exist others constituting the subject of the next section.

\section{Hierarchies from the Third order-First Order Hamiltonian Operators}

The multi-Hamiltonian construction given in this section is based on the following theorem.

{\bf Theorem 1:}{\em Let $u=u(t,x)$ denote the dependent variable of an evolution equation, and $h$ be any arbitrary function of $u$. Then
\begin{eqnarray}
{\mathcal J}=D\frac{1}{u_{x}}D\frac{1}{u_{x}}D+\sqrt{h}D\sqrt{h} \label{huc}
\end{eqnarray}
is a Hamiltonian operator.}

For the {\bf Proof}, one needs to show that ${\mathcal J}$ is skew-adjoint and satisfies the Jacobi identity. The operator ${\mathcal J}$ is manifestly skew-adjoint. The proof of the Jacobi identity is, however, too lengthy to be given in all details here. Therefore, we give only the necessary steps of the proof by the functional multivectors technique \cite{OLV}. According to the technique, a skew-adjoint operator ${\mathcal J}$ satisfies the Jacobi identity if and only if the functional three vector
\begin{eqnarray}
\begin{array}{c}
\Psi=\int\{ \theta \wedge prV_{{\mathcal J}\theta}({\mathcal J}) \wedge \theta\} {\rm d}x, \label{trivec}
\end{array}
\end{eqnarray}
where $prV_{{\mathcal J}\theta}$ is the prolongation of the differential function ${\mathcal J}(\theta)$, vanishes. Starting from the right normal form of the operator (\ref{huc})
\begin{eqnarray}
{\mathcal J}=u_{x}^{-2}D^{3}-3u_{x}^{-3}u_{xx}D^{2}-\
(u_{x}^{-3}u_{xxx}-3u_{x}^{-4}u_{xx}^{2}-h)D+\frac{1}{2}h_{u}u_{x},\label{RNF}
\end{eqnarray}
and using the usual properties of wedge product one obtains
\begin{eqnarray}
\begin{array}{ll}
\Psi=\int \big\{&-u_{x}^{-5}\; \theta \wedge \theta_{x} \wedge \theta_{xxxxxx}\
-3u_{x}^{-5}\; \theta \wedge \theta_{xx} \wedge \theta_{xxxxx}\
-2u_{x}^{-5}\; \theta \wedge \theta_{xxx} \wedge \theta_{xxxx}\\
&+15u_{x}^{-6}u_{xx}\;  \theta \wedge \theta_{x} \wedge \theta_{xxxxx}\
+30u_{x}^{-6}u_{xx}\; \theta \wedge \theta_{xx} \wedge \theta_{xxxx}\\
&+(19u_{x}^{-6}u_{xxx}-102u_{x}^{-7}u_{xx}^{2}-hu_{x}^{-3})\;\theta \wedge \theta_{x} \ \wedge\theta_{xxxx}\\
&+(19u_{x}^{-6}u_{xxx}-102u_{x}^{-7}u_{xx}^{2}-hu_{x}^{-3})\;\theta \wedge \theta_{xx} \wedge \ \theta_{xxx}\\
&+(12u_{x}^{-6}u_{xxxx}-177 u_{x}^{-7} u_{xx}u_{xxx}\\
&\;\;\;\;+6hu_{x}^{-4}u_{xx}+366u_{x}^{-8}u_{xx}^{3}\
+\frac{1}{2}h_{u}u_{x}^{-2})\; \theta \wedge \theta_{x} \wedge \theta_{xxx}\\
&+(3u_{x}^{-6}u_{xxxxx}-57u_{x}^{-7}u_{xx}u_{xxxx}-39u_{x}^{-7}u_{xxx}^{2}\
+489u_{x}^{-8}u_{xx}^{2}u_{xxx}\\
&\;\;\;\;+3hu_{x}^{-4}u_{xxx}\
 -12hu_{x}^{-5}u_{xx}^{2}-576u_{x}^{-9}u_{xx}^{4}+\frac{3}{2}h_{uu}u_{x}^{-1})\; \
\theta \wedge \theta_{x} \wedge \theta_{xx} \big\} {\rm d}x,
\end{array}\label{exptrv}
\end{eqnarray}
for (\ref{trivec}) with (\ref{RNF}). After various integration by parts and evaluation of boundary terms to zero, one obtains the following equalities for any functional $G[u]$ of the dynamical variable $u$
\begin{eqnarray}
\begin{array}{ll}
\int G \theta \wedge \theta_{x} \wedge \theta_{xxxxxx}{\rm d}x=&\int (G \theta \wedge \theta_{xxx} \wedge \theta_{xxxx}-3G^{(1)}\theta_{x} \wedge \theta_{xx} \wedge \theta_{xxx}\\
&\;\;\;-3G^{(2)}\theta \wedge \theta_{xx} \wedge \theta_{xxx}\
+G^{(4)}\theta \wedge \theta_{x} \wedge \theta_{xx}){\rm d}x,
\end{array}\label{eqv1}
\end{eqnarray}
\begin{eqnarray}
\begin{array}{ll}
\int G \theta \wedge \theta_{xx} \wedge \theta_{xxxxx}{\rm d}x=&\int (-G\theta \wedge \theta_{xxx} \wedge\ \theta_{xxxx}+2G^{(1)}\theta_{x} \wedge \theta_{xx} \wedge \theta_{xxx}\\
&\;\;\;+G^{(2)}\theta \wedge \theta_{xx} \wedge \theta_{xxx}){\rm d}x,
\end{array}\label{eqv2}
\end{eqnarray}
\begin{eqnarray}
\begin{array}{ll}
\int G \theta \wedge \theta_{x} \wedge \theta_{xxxxx}{\rm d}x=&\int (G \theta_{x} \wedge \theta_{xx} \wedge \theta_{xxx}+2G^{(1)}\theta \wedge \theta_{xx} \wedge \theta_{xxx}\\
&\;\;\;-G^{(3)}\theta \wedge \theta_{x} \wedge \theta_{xx}){\rm d}x,
\end{array}\label{eqv3}
\end{eqnarray}
\begin{eqnarray}
\begin{array}{ll}
\int G \theta \wedge \theta_{x} \wedge \theta_{xxxx}{\rm d}x=&\int (G^{(2)}\theta \wedge \theta_{x} \ \wedge  \theta_{xx}-G \theta \wedge \theta_{xx} \wedge \theta_{xxx}){\rm d}x,
\end{array}\label{eqv4}
\end{eqnarray}
\begin{eqnarray}
\begin{array}{ll}
\int G \theta \wedge \theta_{x} \wedge \theta_{xxx}{\rm d}x=&\int (-G^{(1)}\theta \wedge \theta_{x} \wedge \theta_{xx}){\rm d}x,
\end{array}\label{eqv5}
\end{eqnarray}
where $G^{(m)}={\rm d}^{m}G/{\rm d}x^{m}$, $m \in {\mathbb {N}}$. When the equalities (\ref{eqv1})-(\ref{eqv5}) are used in (\ref{exptrv}) with the appropriate coefficients of the threevectors in (\ref{exptrv}) replacing $G$'s in (\ref{eqv1})-(\ref{eqv5}), one obtains zero as the coefficients of all the basis trivectors $\theta \wedge \theta_{xxx} \wedge \theta_{xxxx}$, $\theta_{x} \wedge \theta_{xx} \wedge \theta_{xxx}$,
$\theta \wedge \theta_{xx} \wedge \theta_{xxx}$ and $\theta \wedge \theta_{x} \wedge \theta_{xx}$ without any restriction on the function $h$. This proves the theorem.

As a consequence of the above theorem we can consider the following Hamiltonian operators
\begin{equation}
{\mathcal J}_{p}=D\frac{1}{u_{x}}D\frac{1}{u_{x}}D+pDp,\;\;\;\; \
{\mathcal J}_{q}=\frac{1}{q}D\frac{1}{q},
\label{HO34}
\end{equation}
where $p=p(u)$ and $q=q(u) \neq 0$ are arbitrary functions. These operators are compatible since ${\mathcal J}_{p}+\alpha{\mathcal J}_{q}$, $\alpha \in {\mathbb{R}}$, is of form (\ref{huc}) with $h=p^{2}+\alpha /q^{2}$.  The first member (succeeding the trivial symmetry $u_{t_{0}}=0$) of the hierarchy is
\begin{equation}
u_{t_{1}}=(q_{uuu}+p^{2}q_{u}+qpp_{u})u_{x}={\mathcal J}_{p}{\rm E}(H_{-1})=\
{\mathcal J}_{q}{\rm E}(H_{0}),\label{GN2}
\end{equation}
where  $H_{-1}$ ( the Casimir of ${\mathcal J}_{q}$ ) and $H_{0}$ are the first two members of the infinite sequence of Hamiltonian densities

\begin{equation}
\begin{array}{ll}
H_{-1} =& \int q {\rm d}u, \\
H_{0}=& \int (q_{uu} q^{2} - \frac{1}{2} q_{u}^{2} q + \frac{1}{2} q^{3} p^{2}) {\rm d}u \\
H_{1} =& \int \{ q_{uuuu} q^{4} + 2 q_{uuu} q_{u} q^{3} + \frac{3}{2} q_{uu}^{2} q^{3} \
- \frac{3}{2} q_{uu} q_{u}^{2} q^{2} + \frac{5}{2} q_{uu} q^{4} p^{2} \\ 
&+ p_{uu} q^{5} p + \frac{3}{8} q_{u}^{4} q + \frac{5}{4} q_{u}^{2} q^{3} p^{2} \
+ 5 q_{u} p_{u} q^{4} p \
+ p_{u}^{2} q^{5} + \frac{3}{8} q^{5} p^{4} \} {\rm d}u, \\
H_{2} =& \int \{ q_{uuuuuu} q^{6} + 9 q_{uuuuu} q_{u} q^{5} + 12 q_{uuuu} q_{uu} q^{5} \
+ \frac{33}{2}  q_{uuuu} q_{u}^{2} q^{4} + \frac{7}{2} q_{uuuu} q^{6} p^{2}  \\
&+ p_{uuuu} q^{7} p + \frac{9}{2} q_{uuu}^{2} q^{5} + 24 q_{uuu} q_{uu} q_{u} q^{4} \
+ 3 q_{uuu} q_{u}^{3} q^{3} + 21 q_{uuu} q_{u} q^{5} p^{2} \\
&+ 14 q_{uuu} p_{u} q^{6} p + 14 p_{uuu} q_{u} q^{6} p \
+ 4 p_{uuu} p_{u} q^{7} + \frac{5}{2} q_{uu}^{3} q^{4} \
-\frac{15}{4} q_{uu}^{2} q_{u}^{2} q^{3} + \frac{63}{4} q_{uu}^{2} q^{5} p^{2} \\
&+ 21 q_{uu} p_{uu} q^{6} p + \frac{15}{8} q_{uu} q_{u}^{4} q^{2} \
+ \frac{147}{4} q_{uu} q_{u}^{2} q^{4} p^{2} + 105 q_{uu} q_{u} p_{u} q^{5} p \
+ 21 q_{uu} p_{u}^{2} q^{6} \\
&+ \frac{35}{8} q_{uu} q^{6} p^{4} + 3 p_{uu}^{2} q^{7} \
+ \frac{105}{2} p_{uu} q_{u}^{2} q^{5} p + 42 p_{uu} q_{u} p_{u} q^{6} \
+ \frac{5}{2} p_{uu} q^{7} p^{3} - \frac{5}{16} q_{u}^{6} q \
+ \frac{63}{16} q_{u}^{4} q^{3} p^{2} \\
&+ \frac{105}{2} q_{u}^{3} p_{u} q^{4} p \
+ \frac{105}{2} q_{u}^{2} p_{u}^{2} q^{5} + \frac{105}{16} q_{u}^{2} q^{5} p^{4} \
+ \frac{35}{2} q_{u} p_{u} q^{6} p^{3} + 5 p_{u}^{2} q^{7} p^{2} \
+ \frac{5}{16} q^{7} p^{6} \}{\rm d}u, \\
\;\;\;\vdots&
\end{array}\nonumber
\end{equation}
As in the previous section, if we restrict the coefficient of $u_{x}$ in (\ref{GN2}) as
\begin{equation}
q_{uuu}+p^{2}q_{u}+qpp_{u}=u,\nonumber
\end{equation}
and introduce, for convenience, a new arbitrary function $s=s(u)$, $s_{u} \neq 0$, the former arbitrary functions $p$ and $q$ can be written as 
\begin{equation}
p=\sqrt{2\frac{su^{2}}{s_{u}^{\;2}}-2\frac{s_{uuu}}{s_{u}}\
+\frac{s_{uu}^{\;2}}{s_{u}^{\;2}}+2\frac{s_{uu}}{us_{u}}-\frac{3}{u} }\ ,\;\;\;q=\frac{s_{u}}{u},\nonumber
\end{equation}
after the restriction. Let us call the restricted forms of ${\mathcal J}_{p}$ and ${\mathcal J}_{q}$ as ${\mathcal J}_{3}$ and ${\mathcal J}_{4}$ respectively. The final result is 
the following hierarchy which starts from the Casimir $\int{\frac{s_{u}}{u}{\rm d}u}$ of ${\mathcal J}_{4}$.
\begin{eqnarray}
\begin{array}{lll}
u_{t_{0}}&=&0={\mathcal J}_{4}{\rm E}(\int{\frac{s_{u}}{u}{\rm d}u}) \\
u_{t_{1}}&=&uu_{x}={\mathcal J}_{3}{\rm E}(\int{\frac{s_{u}}{u}{\rm d}u})=\
{\mathcal J}_{4}{\rm E}(\int{\frac{ss_{u}}{u}{\rm d}u}) \\
u_{t_{2}}&=&\left[3us+\frac{2}{u}(s_{u}s_{uuu}+2s_{uu}^{2})\
-\frac{7}{u^{2}}s_{u}s_{uu}+\frac{3}{u^{3}}s_{u}^{2}\right]u_{x}\\
&=&{\mathcal J}_{3}{\rm E}(\int{\frac{ss_{u}}{u}{\rm d}u})\
={\mathcal J}_{4}{\rm E}(\int{(\frac{3}{2}\frac{s^{2}s_{u}}{u}+\
2\frac{s_{u}^{3}s_{uu}}{u^{3}}-\frac{s_{u}^{4}}{u^{4}}) {\rm d}u}) \\
&\;\vdots&
\end{array}\nonumber
\end{eqnarray}

\section{Conclusions}

As a first order PDE, integrability of the Riemann equation is not of much concern  since all first order equations are transformable to each other by at most contact transformations.
Moreover, although a bi-Hamiltonian representation is sufficient for the integrability of an equation, there are four distinct Hamiltonian representations of the Riemann equation already given. Therefore, with so many Hamiltonian structures, the Riemann equation was one of the richest of the completely integrable equations.  Now it is even more richer. It has two different infinite-Hamiltonian structures. From the multi-Hamiltonian point of view this is the first example where an equation is written as infinite-Hamiltonian.

We have also shown here that all of the formerly given Hamiltonian operators of the Riemann equation and the ones added in this work are mutually compatible operators. Therefore, among the Riemann hierarchies there is not an exception to the standart multi-Hamiltonian theory which requires compatibility of the Hamiltonian operators. In this respect, we reinvestigate the bi-Hamiltonian structure of the Hirota-Satsuma system which is claimed to be incompatible in \cite{OE}. The result will be published elsewhere.

\begin{acknowledgments}
The author thank to  Prof. M. G{\"u}rses and Prof. A. Karasu  for their continuous interest and comments and also to  Dr. Artemiy V. Kiselev for his suggestions on the manuscript.
\end{acknowledgments}

\end{document}